\documentclass[tightenlines,superscriptaddress,preprint,nofootinbib,amsmath,amssymb,aps,prd]{revtex4-2}
\usepackage{dcolumn}
\usepackage{amsthm, bm, color, graphicx, graphics, float, subfigure, geometry, hyperref, braket, ulem, booktabs, bookmark, multirow}%
\usepackage[mathlines]{lineno}
\usepackage{silence}
\hypersetup{hidelinks}
\newcommand{\wei}[1]{\mathrm{d}#1}

\newcommand{\lrx}[1]{\left(#1\right)}
\newcommand{\lrz}[1]{\left[#1\right]}
\newcommand{\lrd}[1]{\left\{#1\right\}}
\newcommand{\non}{\nonumber\\}
\begin{document}
\title{Di-\texorpdfstring{$\pi^0$}{pi0} Production and Generalized Distribution Amplitudes at Future Electron-Ion Colliders}
\author{Bing'ang Guo}
 \affiliation{%
 Institute of Modern Physics, Chinese Academy of Sciences, Lanzhou 730000, China}
 \affiliation{%
School of Nuclear Science and Technology, University of Chinese Academy of Sciences, Beijing 100049, China}
 \author{Jing Han}%
 \affiliation{School of Physics, Zhengzhou University, Zhengzhou, Henan 450001, China}
 \author{Ya-Ping Xie}%
 \email{xieyaping@impcas.ac.cn}
 \affiliation{%
 Institute of Modern Physics, Chinese Academy of Sciences, Lanzhou 730000, China}
 \affiliation{%
 University of Chinese Academy of Sciences, Beijing 100049, China}
 \author{Xurong Chen} 
 \email{xchen@impcas.ac.cn}
\affiliation{%
 Institute of Modern Physics, Chinese Academy of Sciences, Lanzhou 730000, China}
 \affiliation{%
School of Nuclear Science and Technology, University of Chinese Academy of Sciences, Beijing 100049, China}
\affiliation{%
Southern Center for Nuclear Science Theory (SCNT), Institute of Modern Physics,
 Chinese Academy of Sciences, Huizhou 516000, Guangdong Province, China}
\begin{abstract}
Generalized distribution amplitudes (GDAs) offer valuable insights into the three-dimensional structure of hadrons, delineating the amplitudes associated with the transition from a quark-antiquark pair to a hadron pair. Currently, hadron GDAs can be probed in electron-positron collisions, with experimental feasibility demonstrated at facilities such as Belle and BESIII. In this study, we put forth the proposition that hadron GDAs can also be investigated in electron-hadron collisions at forthcoming Electron-Ion Colliders (EICs), specifically through the subprocess $\gamma^*\gamma \to h_1 h_2$. In this framework, a quasi-real photon, emitted by the ion, exhibits a photon flux proportional to the square of the ion's electric charge. Consequently, we anticipate that the cross sections in EICs will be substantially larger than those in electron-positron collisions. We present numerical calculations pertaining to di-$\pi^0$ production employing the equivalent photon approximation (EPA). Our findings suggest that, within the same kinematic region, electron-proton ($e$-$p$) collisions at the EIC could yield an event rate comparable to that of Belle II, while electron-gold ($e$-Au) collisions are expected to generate an even greater number of events. This enhanced event rate facilitates a high-precision examination of di-$\pi^0$ GDAs at the EIC.
\end{abstract}
\maketitle
\newpage
\section{\label{int}introduction}

Generalized distribution amplitudes (GDAs) have been introduced as a framework for investigating the three-dimensional structure functions of hadrons \cite{Muller:1994ses, Diehl:1998dk, Diehl:2000uv}. These amplitudes can be studied in time-like processes, such as the production of hadron-antihadron pairs via di-photon interactions (\( \gamma^* \gamma \rightarrow h \bar{h} \)) \cite{Diehl:2000uv} and in electron-positron annihilation events (\( e^+ e^- \rightarrow h \bar{h} \gamma \)) \cite{Lu:2006ut, Pire:2023kng, Pire:2023ztb}. Additionally, GDAs are relevant for the examination of energy-momentum tensor (EMT) form factors \cite{Polyakov:2018zvc, Burkert:2023wzr, Muller:1994ses, Diehl:1998dk, Polyakov:1998ze, Goeke:2001tz, Kumano:2017lhr}. They also significantly contribute to our understanding of the decay processes of \( B \) mesons \cite{Chen:2002th, Wang:2015uea, Li:2016tpn}. Moreover, GDAs can yield insights into exotic states, including both diffuse molecular configurations and compact multiquark structures \cite{Kawamura:2013wfa, Kawamura:2013iia, Chang:2015ioc}. Furthermore, GDAs can be utilized to investigate the gravitational radius of hadrons.

In particular, GDAs for the \( \pi^0 \) meson have been analyzed in the interaction process \( e \gamma \rightarrow e \pi^0 \pi^0 \) by the Belle collaboration \cite{Kumano:2017lhr}. Recent studies suggest that kinematic higher-twist effects may influence the cross sections measured at BESIII and Belle in specific kinematic regions \cite{Lorce:2022tiq, Pire:2023kng, Pire:2023ztb}.

The forthcoming Electron-Ion Collider (EIC) and the Electron-Ion Collider of China (EicC) are specifically designed to investigate the structure of hadrons across various kinematic regions \cite{Accardi:2012qut, Anderle:2021wcy}. In the context of electron-proton ($e$-$p$) and electron-nucleus ($e$-$A$) collisions, incident particles interact through the exchange of photons, which subsequently facilitates the production of di-$\pi^0$ pairs. Consequently, the process \( \gamma^*\gamma \rightarrow \pi^0\pi^0 \) can be analyzed within $e$-$p$ and $e$-$A$ collisions. Therefore, electron-hadron collisions serve as a valuable avenue for extracting $\pi^0$ GDAs through the exclusive production of di-$\pi^0$s at EICs. Additionally, the upgraded Belle II experiment is expected to achieve a significantly higher luminosity, making it a viable platform for the study of hadron GDAs as well.

In this study, we examine the exclusive di-$\pi^0$ production in both electron-hadron and \( e^+\)-\(e^- \) collision scenarios. In the di-photon process, the photon flux emitted by the hadron plays a pivotal role in determining the differential cross section. This photon flux is employed to quantify the density of quasi-real photons produced by hadrons and positrons, utilizing the equivalent photon approximation (EPA) method to calculate the di-$\pi^0$ production cross sections. The EPA has been extensively applied in the analysis of particle photoproduction processes, particularly in the context of ultra-peripheral collisions (UPCs) \cite{Bertulani:2005ru, Xie:2018rog}. For the transition from di-photon to di-$\pi^0$ production, GDAs will be incorporated into the computational framework.

The paper is organized as follow. Theoretical frame is presented in Sec.~\ref{for}.  The numerical results.~\
are exhibited in Sec.~\ref{res}. Summaries and conclusions are given in Sec.~\ref{sum}. 

\section{\label{for}Theoretical framework}

\begin{figure}[!ht]
  \centering
  \subfigure{
    \includegraphics[width=0.36\textwidth]{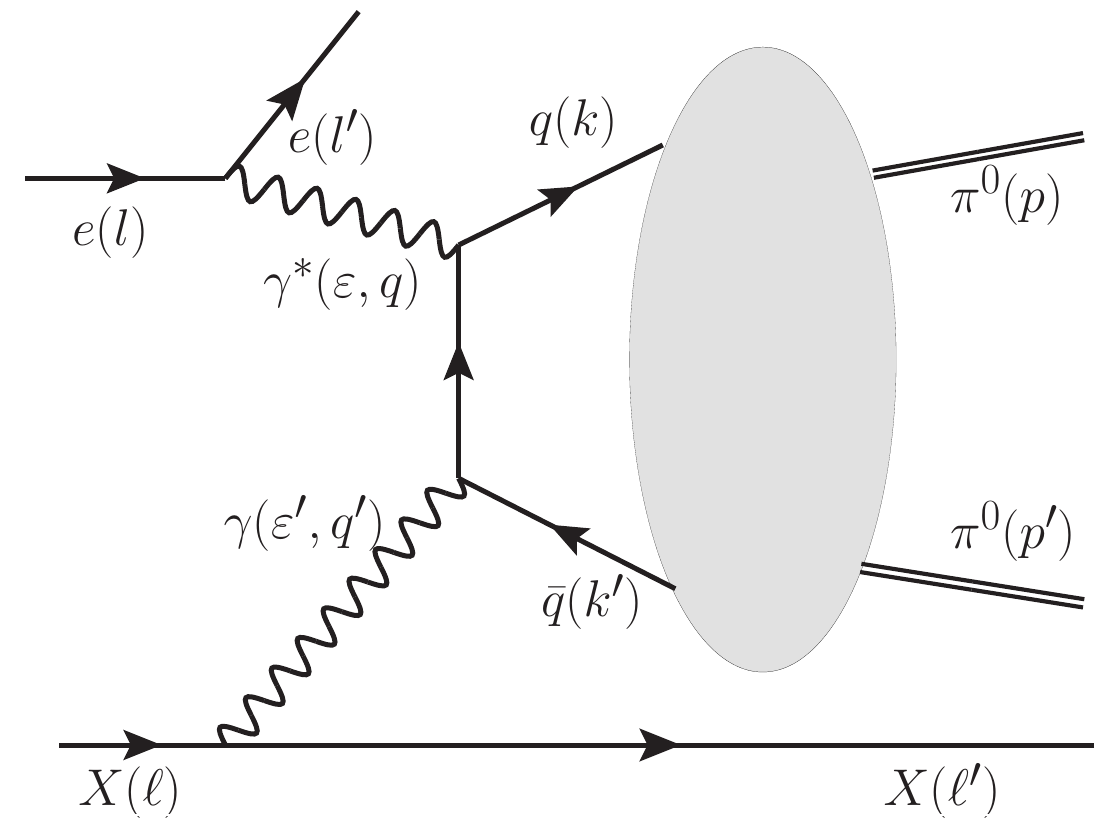}
  }
  \subfigure{
    \includegraphics[width=0.36\textwidth]{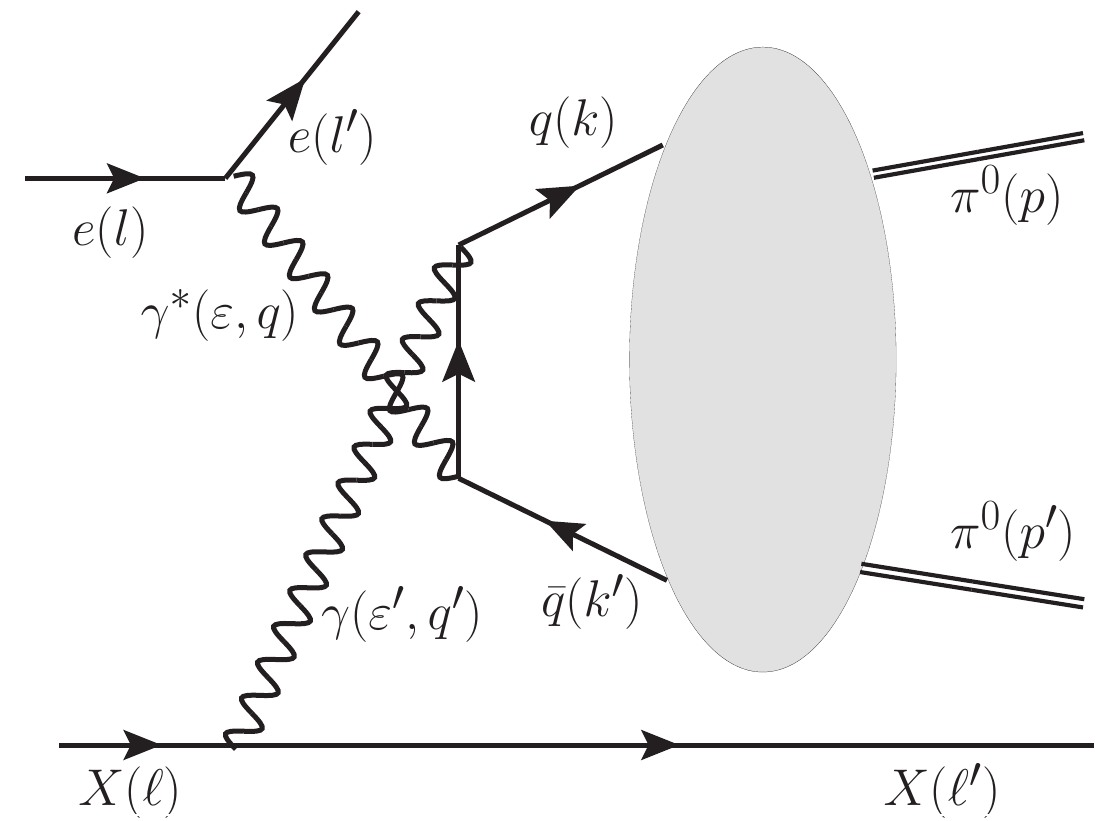}
  }
  \caption{Feynman diagram for di-$\pi^0$ production in $e$-$X$ collisions via quark GDAs (left panel), with the right panel showing the corresponding crossed figure of the same process..}
  \label{gda}
\end{figure}

In electron-ion collisions at the EICs, the interaction between electrons and hadrons occurs via the exchange of photons. Specifically, hadrons can emit quasi-real photons, while electrons, in the high-energy limit, can emit virtual photons. These virtual photons may interact with quasi-real photons to produce quark-antiquark ($q\bar{q}$) pairs or gluon ($gg$) pairs. The nonperturbative amplitudes associated with the combination of $q\bar{q}$ pairs into di-hadrons can be parametrized using quark GDAs, while the process of hadron pairs via $gg$ pairs is described by gluon GDA. The relevant Feynman diagrams for the $q\bar{q}$ production process are illustrated in Fig.~\ref{gda}. For the purpose of this analysis, we neglect contributions from the gluon GDA, as these are suppressed by higher-order terms of the running coupling constant, $\alpha_s$.

\begin{figure}[!ht]
    \centering
    \includegraphics[width=0.6\textwidth]{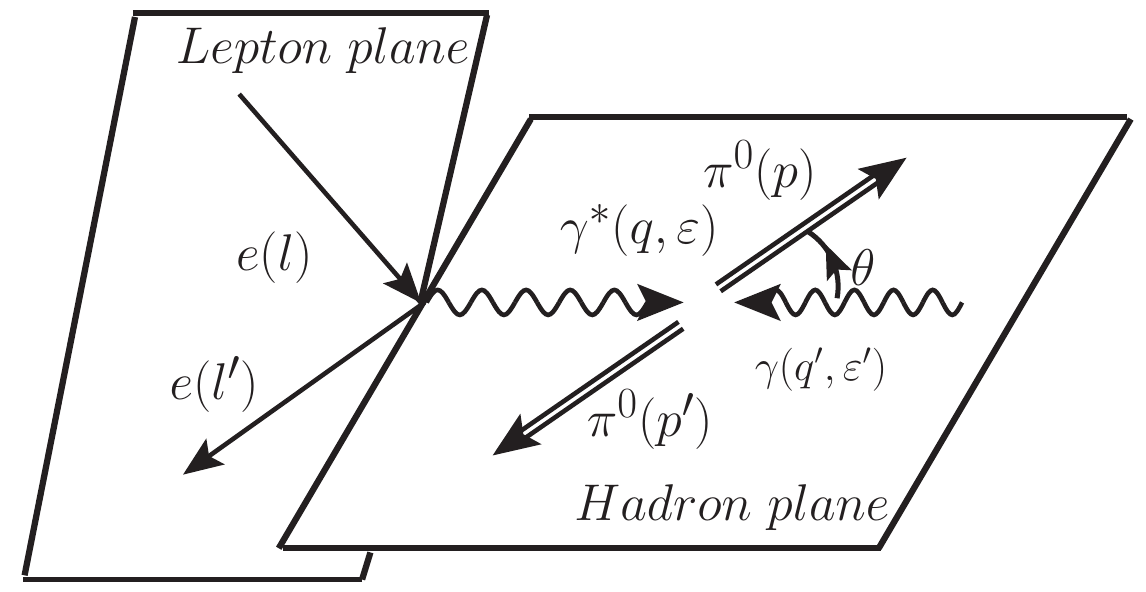}
    \caption{$e\gamma\rightarrow e\pi^0\pi^0$ process in the $\gamma^*\gamma$ center-of-mass frame.}
    \label{frame}
\end{figure}

Fig.~\ref{frame} shows the $e\gamma\rightarrow e\pi^0\pi^0$ process in the c.m. frame of the di-photon, corresponding momentum are assigned as:
\begin{eqnarray}
  &q=(q^0,0,0,|\vec{q}|),~q'=(|\vec{q}|,0,0,-|\vec{q}|),&\non
  &p=(p^0,|\vec{p}|\sin\theta,0,|\vec{p}|\cos\theta),~p'=(p^0,-|\vec{p}|\sin\theta,0,-|\vec{p}|\cos\theta).&
\end{eqnarray}
Three unit polarization vectors are set as
\begin{equation}
  \varepsilon_0=\frac{1}{Q}(|\vec{q}|,0,0,q^0),~\varepsilon_{\pm}=\frac{1}{\sqrt{2}}(0,\mp 1,-i,0),~\varepsilon^\prime_\pm=\frac{1}{\sqrt{2}}(0,\mp 1,i,0),
\end{equation}
indices 0, $\pm$ denote the helicity of the photon. Quasi-real photon is only transversely polarized while the virtual one has longitudinal component.

In the framework of EPA, the cross section of $eX\rightarrow eX\pi^0\pi^0$ can be expressed as:
\begin{equation}
  \frac{\mathrm{d}\sigma_{eX\rightarrow eX\pi^0\pi^0}}{\wei{Q^2}\wei{W^2}}
  =\int^{1}_{0}\wei{x}
  \frac{\wei{\sigma}_{e\gamma\rightarrow e\pi^0\pi^0}}{\wei{Q^2}\wei{W^2}}
  f_{\gamma/X}(x)\delta\lrx{x-\frac{s_{e\gamma}}{s_{eX}}},
\end{equation}
where $W^2=P^2=(q+q')^2=(p+p')^2\equiv s_{\pi^0\pi^0}$ is the invariant-mass squared of the $\pi^0$ pair, $Q^2=-q^2$, $x=q\cdot q^\prime/q\cdot\ell$ is ratio between the energy of quasi-real photon and the source particle. Photon flux can be understood as the possibility for a certain incident particle to emit a photon with energy $E_\gamma=xE_{X}$ and is defined by the Fourier transformation of the spatial distribution of electromagnetic fields surrounding the incident particle. Photon flux for a proton is given as \cite{Xie:2018rog,Bertulani:2005ru}:
\begin{equation}
  f_{\gamma/p}(x)=\frac{\alpha_{\mathrm{em}}}{2\pi x}\lrz{1+\lrx{1-x}^2}\lrx{\ln\Omega-\frac{11}{6}+\frac{3}{\Omega}-\frac{3}{2\Omega^2}+\frac{1}{3\Omega^3}},
\end{equation}
where $\Omega=1+0.71$ GeV$^2/Q^2_{\mathrm{min}}$, $Q^2_{\mathrm{min}}=x^2E_p^2/\gamma^2_L$, $\gamma_L=E_p/m_p$ is the Lorentz factor of proton. 

While in the case of nucleus with charge number $Z>1$, more nucleons will contribute to the emission of photon, the photon flux is written as \cite{Xie:2018rog,Bertulani:2005ru}:
\begin{equation}
  f_{\gamma/A}(x)=\frac{2Z^2\alpha_{\mathrm{em}}}{\pi x}\lrd{\xi K_0(\xi)K_1(\xi)-\frac{\xi^2}{2}\lrz{K^2_1(\xi)-K^2_0(\xi)}},
\end{equation}
where $K_0(\xi)$ and $K_1(\xi)$ are modified Bessel functions and with $\xi=2x R_A m_p$ and $R_A=(1.12A^{1/3}-0.86A^{-1/3})$ fm is the charge radius of the nucleus \cite{Wang:2023zay,Lappi:2010dd}.

Moreover, we also consider the di-$\pi^0$ production in $e^+$-$e^-$ collision at Belle II. The quasi-real photon flux for electron/positron employed in the calculation of differential cross section of process $e^+e^-\lrx{\gamma^*\gamma}\rightarrow e^+e^-\pi^0\pi^0$ is presented as \cite{Berger:1986ii}:
\begin{equation}
    f_{\gamma/e^+}(x)=\frac{\alpha_{\mathrm{em}}}{\pi x}\lrd{\lrz{1+(1-x)^2}\ln\lrz{\frac{\gamma_L(1-x)\theta_{\mathrm{max}}}{x}}-1+x},
    \label{fluxe}
\end{equation}
where $\theta_{\mathrm{max}}$ is the maximum allowed angle for the photon to be regarded as quasi-real, referring to the Belle II technical design report \cite{Belle-II:2010dht}, we obtain $\theta_{\mathrm{max}}=17^\circ$.

The next step,  we consider the $e\gamma\rightarrow e\pi^0\pi^0$ cross section. According to Refs.~\cite{Diehl:2000uv,Lorce:2022tiq}, differential cross section of the $e\gamma$ subprocess is given as
\begin{eqnarray}
  & &\frac{\wei{\sigma}_{e\gamma\rightarrow e\pi^0\pi^0}}{\wei{Q^2}\wei{W^2}\wei{(\cos\theta)}}=\frac{\alpha_{\mathrm{em}}^3 \beta}
  {8s_{e\gamma}^2} \frac{1}{Q^2(1-\epsilon)} \non
  & &\qquad\qquad\qquad\qquad
  \times\left[\left|A_{++}\right|^2+\left|A_{-+}\right|^2+2 \epsilon\left|A_{0+}\right|^2\right] ,
  \label{sub}
  \end{eqnarray}
  where $\beta=|\vec{p}|/p^0=\sqrt{1-4m^2_{\pi}/W^2}$ is speed of the $\pi^0$, $\theta$ is the angle between directions of the $\pi^0$s and the photons, $\beta$ and $\theta$ are both measured in the c.m. frame of the di-$\pi^0$.
The light-cone momentum fraction for the virtual photon in the incident electron is denoted as:
 $ y=(q\cdot q')/(l\cdot q')=(Q^2+W^2)/s_{e\gamma}$,
  and the ratio of longitudinal and transverse polarization  $\epsilon=(1-y)/(1-y+y^2/2)$. 
  
  And helicity amplitudes in Eq.~(\ref{sub}) are defined as $A_{ij}=\varepsilon^\mu_i \varepsilon_j^{\prime \nu} \mathcal{T}_{\mu\nu}$, where $\mathcal{T}^{\mu\nu}$ is the hadronic tensor:
\begin{eqnarray}
  \mathcal{T}^{\mu\nu}
  &=&i\int\wei{^4\mathrm{y}}\mathrm{e}^{-iq\cdot \mathrm{y}}\braket{\pi(p)\pi(p')|\text{T} \{J^\mu_{\mathrm{em}}(y)J^\nu_{\mathrm{em}}(0)\}|0}.
\end{eqnarray}
Thanks to parity invariance, there are only three independent amplitudes,
\begin{eqnarray}
  & &A_{++}=A_{--}=\mathcal{A}^{(0)},\non
  & &A_{0+}=A_{0-}=-\mathcal{A}^{(1)}(\Delta\cdot\varepsilon_-)=-\frac{\vert\Delta_T\vert}{\sqrt{2}}\mathcal{A}^{(1)},\non
  & &A_{-+}=A_{+-}=-\mathcal{A}^{(2)}(\Delta\cdot\varepsilon_-)^2=-\frac{\vert\Delta_T\vert^2}{2}\mathcal{A}^{(2)}.
\end{eqnarray}
$\Delta$ is the difference between the momentums of the two $\pi^0$s, and its transverse part is $\Delta_T=p^\prime-p-(1-2\zeta)\lrz{\tilde{n}-W^2/(Q^2+W^2)n}$, $\tilde{n}=q+Q^2/(W^2+Q^2)q^\prime,~n=q^\prime$ are two light-like vectors. 
The helicity amplitudes are given in terms of the leading-twist $\pi^0$ GDAs \cite{Lorce:2022tiq}. And in this work, we calculate the cross sections with $\pi^0$ GDAs extracted from leading-twist analysis of the Belle measurements \cite{Kumano:2017lhr}.

  \section{\label{res}numerical results}
  In this section, we  will present the numerical cross section prediction and estimate the number of events for $eX\rightarrow eX\pi^0\pi^0$ in the designed kinematic regions of different facilities using the framework of EPA and GDAs.
    
 \begin{figure}[!ht]
     \centering
     \includegraphics[width=0.5\textwidth]{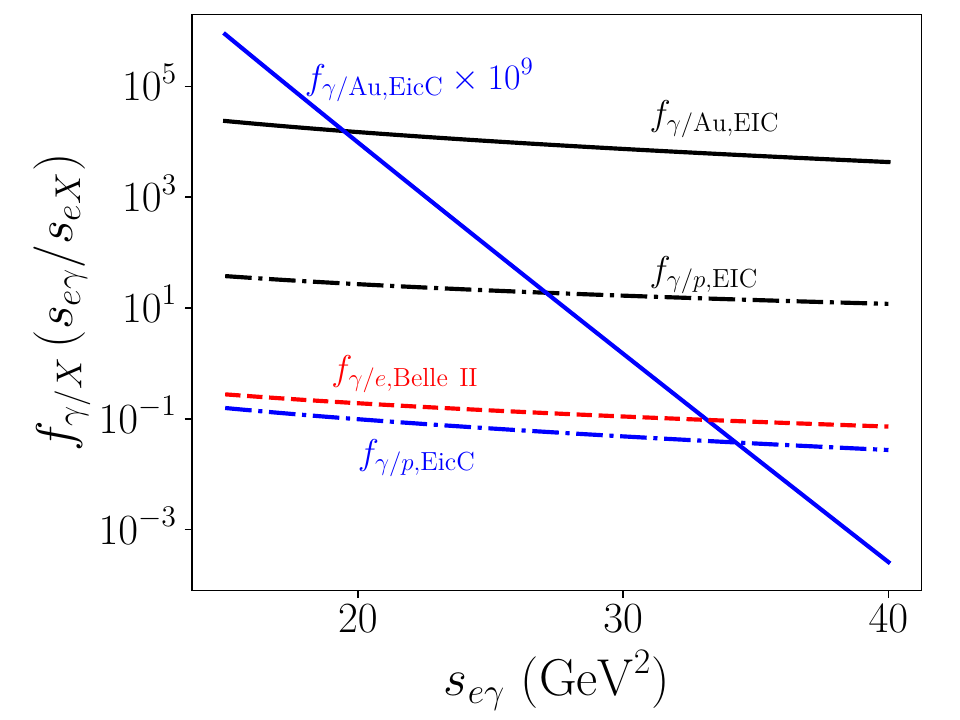}
     \caption{Photon fluxes under different facilities' kinematics as functions of $s_{e\gamma}$.}
     \label{fluxes}
 \end{figure}
Firstly, we focus on the $\gamma/X$ flux in the facilities. Based on the design specifications of EIC, EicC and Belle II \cite{AbdulKhalek:2021gbh,Anderle:2021wcy,Belle-II:2010dht}, the photon flux of hadron and positron can be determined numerically. The different photon fluxes as functions of the c.m. energy of $e\gamma$ system are illustrated in Fig.~\ref{fluxes}. Since the beam energy at EicC is significantly lower than that of the EIC, the ratio $x\equiv s_{eh}/s_{e\gamma}$ at EicC is also much smaller than at EIC. Due to the rapid decrease in the photon fluxes, they will be relatively lower at EicC compared to EIC.  
\begin{figure}[!ht]
  \centering
  \subfigure{
    \includegraphics[width=0.47\textwidth]{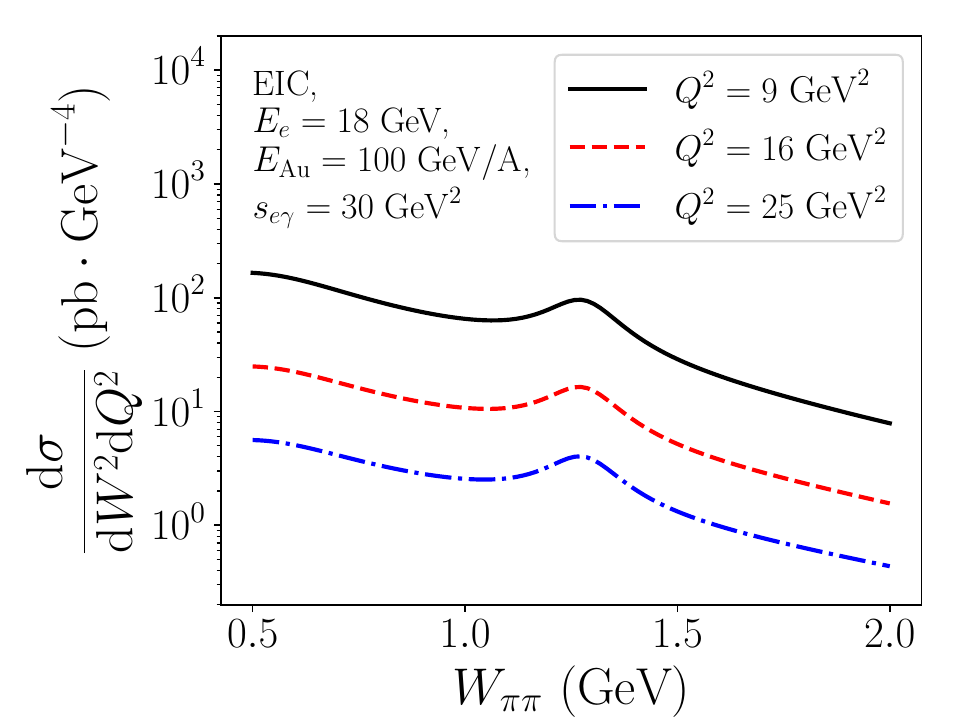}
  }
  \subfigure{
    \includegraphics[width=0.47\textwidth]{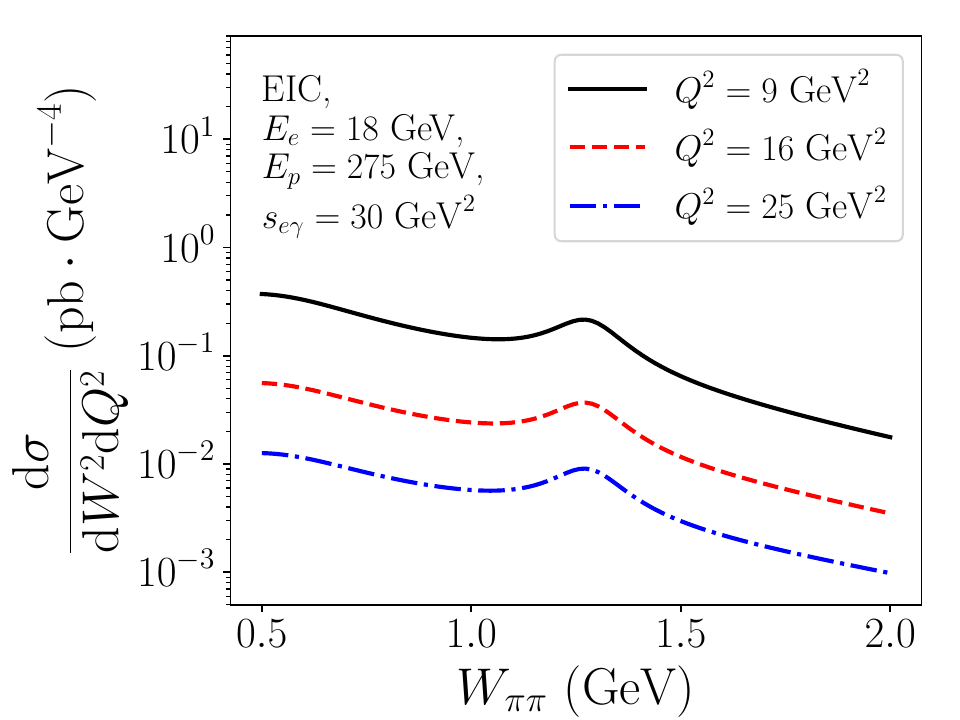}
  }
  \caption{Differential cross sections for di-$\pi^0$ production in $e$-Au (left panel) and $e$-$p$ (right panel) collisions under the designed conditions of EIC within the kinematic region $s_{e\gamma}=30~\mathrm{GeV}^2$, 0.5 GeV$\le W\le$2.0 GeV and $Q^2$=9, 16, 25 GeV$^2$, respectively.}
  \label{fig0}
\end{figure}

\begin{figure}[!ht]
  \centering
  \subfigure{
    \includegraphics[width=0.47\linewidth]{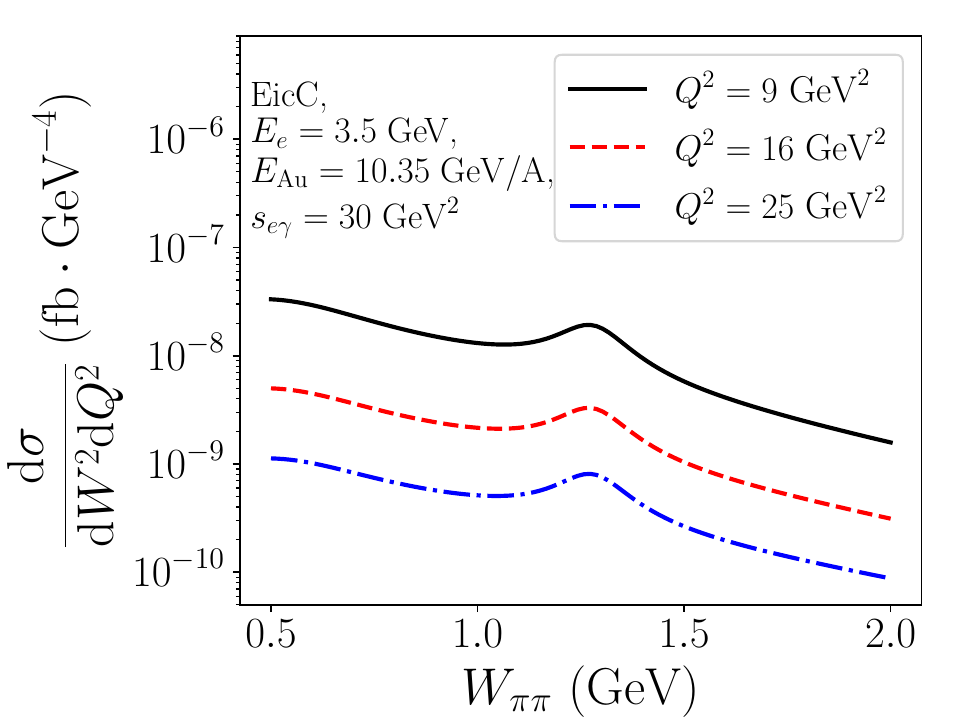}
  }
  \subfigure{
    \includegraphics[width=0.47\linewidth]{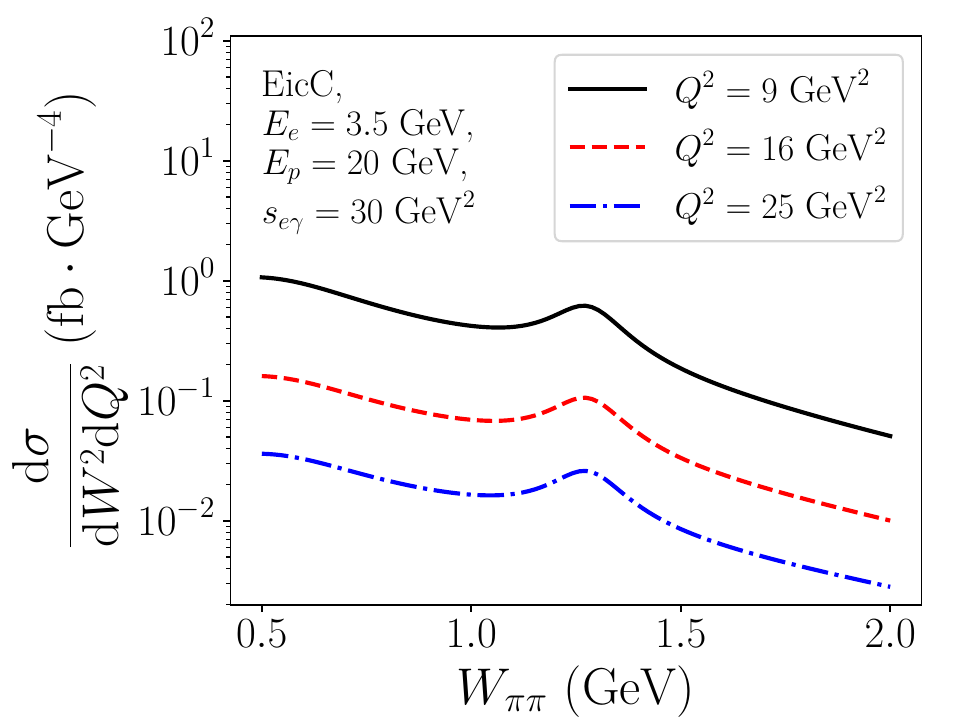}
  }
  \caption{Differential cross sections for di-$\pi^0$ production in $e$-Au (panel (a)) and $e$-$p$ (panel (b)) collisions under the designed conditions of EicC within the kinematic region $s_{e\gamma}=30~\mathrm{GeV}^2$, 0.5 GeV$\le W\le$2.0 GeV and $Q^2$=9, 16, 25 GeV$^2$, respectively.}
  \label{figw2}
\end{figure}

\begin{figure}[!ht]
  \centering
  \subfigure{
  \includegraphics[width=0.47\textwidth]{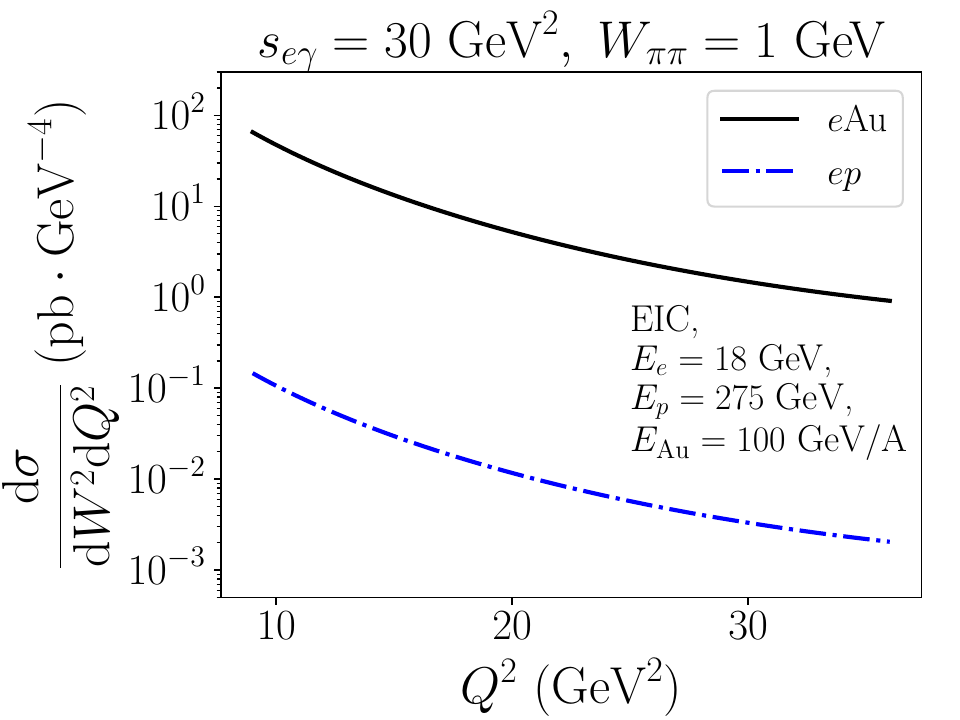}
  }
  \subfigure{
  \includegraphics[width=0.47\textwidth]{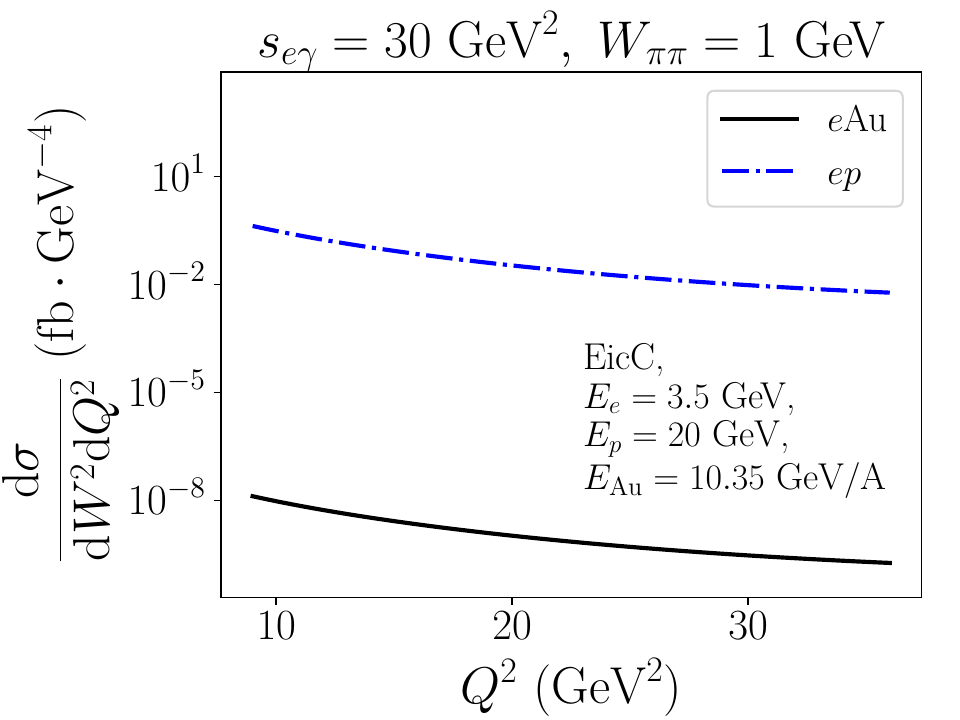}
  }
  \caption{$Q^2$ dependence of the differential cross sections of di-$\pi^0$ production in $e$-$h$ collisions at EIC (left panel) and EicC (right panel), with $s_{e\gamma}=30~\mathrm{GeV}^2,~W_{\pi\pi}=1.0~\mathrm{GeV}$.}
  \label{figq2}
\end{figure}
\begin{figure}[!ht]
  \centering
  \subfigure{
  \includegraphics[width=0.47\textwidth]{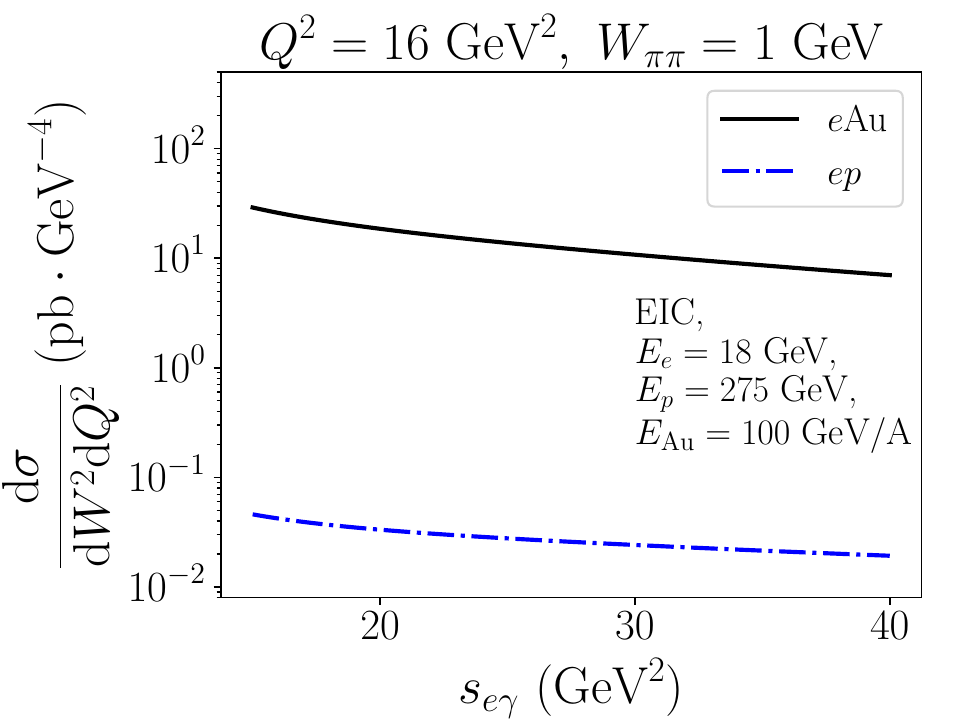}
  }
  \subfigure{
  \includegraphics[width=0.47\textwidth]{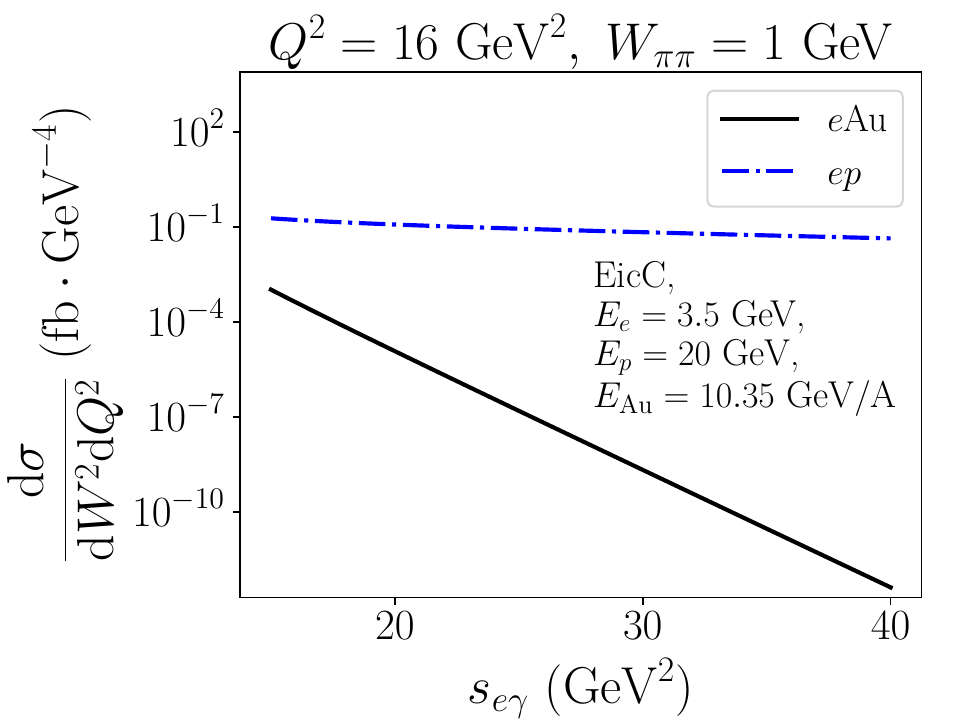}
  }
  \caption{$s_{e\gamma}$ dependence of the differential cross sections of di-$\pi^0$ production in $e$-$h$ collisions at EIC (left panel) and EicC (right panel) with $Q^2=16~\mathrm{GeV}^2,~W_{\pi\pi}=1.0~\mathrm{GeV}$.}
  \label{figs}
\end{figure}

Subsequently, we integrate the photon fluxes, denoted as $\gamma/X$, with the differential cross sections for electron-photon ($e\gamma$) interactions to assess the cross section for the process \( eX \rightarrow eX\pi^0\pi^0 \). The differential cross sections for the process \( eh \rightarrow eh\pi^0\pi^0 \) are presented in Figs. \ref{fig0}, \ref{figw2}, \ref{figq2}, and \ref{figs}. Within the invariant mass range of \( 0.5 \, \text{GeV} \leq W \leq 2.0 \, \text{GeV} \) and for momentum transfer squared values \( Q^2 = 9, 16, 25 \, \text{GeV}^2 \), a pronounced decrease in the cross section as a function of invariant mass is observed, while the dependence on \( Q^2 \) appears to be comparatively weak. Additionally, Fig.~\ref{figs} illustrates the significant influence of the center-of-mass energy \( s_{e\gamma} \) on the differential cross sections.

Furthermore, we estimate the number of di-$\pi^0$ production events across various experimental facilities. For all studies conducted at the EIC, EicC and Belle II, we adopt an overarching detection efficiency of 70\%. According to the Belle II design report, it is anticipated that the facility will accumulate an integrated luminosity of \( 50 \, \text{ab}^{-1} \) over the upcoming decade \cite{Aihara:2024zds}, leading to an average of \( 5 \, \text{ab}^{-1}\cdot\text{y}^{-1} \). Analysis of the results presented in the aforementioned figures reveals that the cross section for di-$\pi^0$ production in $e$-Au interactions at EicC is exceptionally small due to the low c.m. energy available. Consequently, we do not include discussion of the $e$-Au events in the context of the EicC.

The earlier data obtained from the Belle collaboration were characterized by considerable statistical uncertainty; however, following recent upgrades, Belle II is now equipped to measure $\pi^0$ GDAs with enhanced precision. 
With an integrated luminosity of \( 15 \, \text{fb}^{-1}\cdot\text{y}^{-1}\), $e$-$p$ collisions at the EIC are anticipated to produce a number of events comparable to those generated by Belle II, while electron-gold ($e$-Au) collisions are expected to yield an even greater event count (about $10^4$ times). This suggests that the exploration of $\pi^0$ GDAs at the EIC can be undertaken with a higher degree of accuracy.

\section{\label{sum}Conclusions}

The EPA posits that the photon flux generated by heavy ions substantially influences photon-induced processes. UPCs involving heavy ions are particularly advantageous for enhancing observables in high-energy physics, emphasizing the importance of exploiting this characteristic of high-energy heavy ion beams. Therefore, investigating photon-induced processes in electron-ion collisions is crucial for advancing our understanding of the microscopic structure of particles.

In this study, we compute the exclusive differential cross sections for di-$\pi^0$ production in $e$-$p$ and $e$-$A$ collisions using di-$\pi^0$ GDAs as input. Our analysis indicates a decreasing trend in the differential cross sections of di-$\pi^0$s with increasing momentum transfer squared ($Q^2$), while larger cross sections are observed at lower values of the invariant mass ($W$), particularly showcasing a peak associated with the $f_2(1270)$ resonance. The increased electric charge number ($Z$) in $e$-$A$ collisions leads to significantly larger cross sections compared to $e$-$p$ collisions at the EIC. 

These findings are essential for simulating di-$\pi^0$ production processes in both $e$-$p$ and $e$-$A$ collisions and will aid in accurately estimating production cross sections at future EICs. The upcoming electron-ion collisions at the EIC present a promising avenue for investigating GDAs. Given the anticipated kinematic configurations and luminosities across various facilities, interactions between electrons and heavier nuclei at the EIC are expected to yield a greater number of events, along with improved precision. Consequently, we advocate for the exploration of GDAs through electron-heavy ion UPCs at the EIC, as this approach is poised to enhance our understanding of their intrinsic nature. Continued research on GDAs is vital for obtaining deeper insights into the nonperturbative domain of quantum chromodynamics (QCD).

\section{\label{ack}acknowledgement}

The authors are grateful to Prof.~Yanbing Cai for helpful discussion about the physics of photon flux. This work has been supported by the National Key R\&D Program of China (Grant NO. 2024YFE0109800 and 2024YFE0109802).

\bibliography{reference}

\end{document}